\documentclass{article}
\usepackage{graphicx}
\usepackage{bm}
\newlength{\Tatescale}
\newlength{\figwidth}
\newcommand{\Cut}[1]{}
 
\newcommand{\beq}{\begin{eqnarray}}
\newcommand{\eeq}{\end{eqnarray}}

\newcommand{\bp}{\mbox{\boldmath $p$}}

\newcommand{\br}{\mbox{\boldmath $r$}}
\newcommand{\bx}{\mbox{\boldmath $x$}}
\newcommand{\by}{\mbox{\boldmath $y$}}

\newcommand{\bv}{\mbox{\boldmath $v$}}
\newcommand{\bL}{\mbox{\boldmath $L$}}

\newcommand{\bS}{\mbox{\boldmath $S$}}

\newcommand{\bsigma}{\mbox{\boldmath $\sigma$}}

\newcommand{\pslash}{p\kern-1ex /}
\newcommand{\kslash}{k\kern-1ex /}
\newcommand{\qslash}{q\kern-1ex /}
\newcommand{\lslash}{l\kern-1ex /}
\newcommand{\sslash}{s\kern-1ex /}
\newcommand{\paslash}{p_a\kern-2ex /}
\newcommand{\pbslash}{p_b\kern-2ex /}
\newcommand{\Dslash}{{\cal D}\kern-1.5ex /}
\newcommand{\dslash}{\partial\kern-1.2ex /}

\begin{document}

\begin{center}
{\Large{\bf Lattice Nuclear Force}}\footnote{To appear in Festschrift for Gerry Brown, 
ed. Sabine Lee (World Scientific, Singapore)}

\bigskip 

{\large T.\ Hatsuda}\\

\bigskip

{\it Department of Physics, The University of Tokyo, Tokyo 113-0033, Japan}\\
{\it and}\\
{\it IPMU, The University of Tokyo, Chiba, 277-8583, Japan}

\end{center}

\bigskip

\begin{abstract}
We derive an energy-independent and  non-local potential for 
the  baryon-baryon interaction from 
 the Nambu-Bethe-Salpeter amplitude on the lattice.
 The method is applied to the nucleon-nucleon
 interactions with the   (2+1)-flavor lattice QCD simulations.
  The central and tensor potentials are obtained as leading order
   terms of the velocity expansion of the non-local potentialF
 The central potential has a repulsive core surrounded by an
 attractive well, while the tensor potential has intermediate and
  long range attractions. Validity of the velocity expansion
   is tested by the nucleon-nucleon interaction with finite 
    relative momentum.   Interactions among octet-baryons
    in the flavor SU(3) limit are also studied in full QCD
    simulations to unravel the origin of the repulsive core and the
     possible existence of $H$-dibaryon.
\end{abstract}

\section{Introduction}
\label{sec:intro}

Understanding of  the nuclear force from  quantum chromodynamics (QCD)
 is one of the most challenging problems in nuclear and particle physics.
 Experimentally, a large number of proton-proton and neutron-proton
  scattering data as well as deuteron properties 
  have been  accumulated and 
  summarized e.g. in the Nijmegen database \cite{Nijmegen_data}.
 Below the pion production threshold,
  the notion of the $NN$ potential (either in the coordinate space or in
  the momentum space) is useful in the sense that it can be 
   used not only to describe the two-body system but also to 
   study the nuclear many-body problems through ab-initio calculations \cite{BK2010}.

 The phenomenological  $NN$  potentials in the coordinate 
 space are known to reflect some characteristic features of the $NN$  interaction
  \cite{NN-review}:\\
(i)
  The long range part of the nuclear force  (the relative distance 
 $r >  2$  fm) 
 is dominated by the one-pion exchange \cite{yukawa}.
  Because of the pion's Nambu-Goldstone character, it
  couples to the spin-isospin density of the nucleon and hence
  leads to the tensor force.\\
 (ii)  The medium range part ($1\ {\rm fm} < r < 2$ fm) receives 
  significant contributions from the exchange of  
  multi-pions.
  In particular, the spin-isospin independent attraction
 of about 50 -- 100 MeV in this region plays an essential role
  to nuclear binding.\\
(iii) The short  range part ($r < 1$ fm) is best described by
  a strong repulsive core  \cite{jastrow,nambu57}.
  Such a short range repulsion is relevant  
 for the maximum mass of neutron stars.\\
(iv) There is also a strong attractive spin-orbit force in the 
  isospin 1 channel at medium and
 short distances. This leads to the $^3 {\rm P}_2$ neutron pairing
 in neutron matter and hence the neutron
  superfluidity inside neutron stars \cite{VJ}.

 Several high precision $NN$ forces are now available 
  to fit neutron-proton and proton-proton scattering data (about 4500 data points)
   with $\chi^2/{\rm dof} \sim 1$. However, 
     they have typically 20-40 fitting parameters:
   e.g. CD Bonn potential, AV18 potential and N$^3$LO chiral effective field 
   theory have 38, 40, and 24 parameters, respectively \cite{Machleidt:2007ms}.
  If one tries to extend these to hyperon-nucleon and hyperon-hyperon interactions,
  the task becomes extremely tough since  the number of parameters 
   increase and the scattering data are scarce.
       In this situation, it is highly desirable to
  study the general baryon-baryon interactions from the first principle
  lattice QCD simulations, since all the hadronic interactions in QCD are
   controlled only by the
    QCD scale parameter ($\Lambda_{\rm QCD}$) and 
    the quark masses ($m_u$, $m_d$, $m_s$)  whose values are
    pretty well determined \cite{Colangelo:2010et}.

 A theoretical framework to study the hadron-hadron interaction
  using lattice QCD  was first proposed 
  by L\"{u}scher \cite{luescher} and was applied to the lattice  
   simulations for the $NN$ interaction in \cite{Fukugita:1994ve}: For two hadrons in a finite
  box with a size $L \times L \times L$ in the periodic boundary condition,  
  an exact relation between  the energy spectra in the box
  and the elastic scattering phase shift at these energies was
   derived.  If the range of the hadronic interaction $R$  is sufficiently
  smaller than the size of the box $R<L/2$, the behavior of the 
  equal-time Nambu-Bethe-Salpeter (NBS)
  amplitude $\psi (\br)$ in the interval $R < | \br | < L/2 $
  under the periodic boundary condition
  has sufficient information to relate the phase shift and the 
  two-particle spectrum. 
   This  L\"{u}scher's method bypasses
  the difficulty to treat the real-time scattering process
  on the Euclidean   lattice.
  
   Recently, an alternative
  approach to the hadron interactions in lattice QCD
  was proposed  \cite{Ishii:2006ec,Aoki:2008hh,Aoki:2009ji}.  
  The starting point is the
  same equal-time NBS amplitude  $\psi (\br)$:  
  Instead of looking at the amplitude 
  outside the range of the interaction,
   the internal region $ |\br | < R$ is considered and
  an energy-independent  non-local potential $U(\br, \br')$
  is defined from  $\psi (\br)$.
  Since $U(\br, \br')$ in QCD
   is a localized function in space  due to confinement
   of quarks and gluons, it receives
   finite volume effect only weakly. Therefore, 
   once $U$, although it is not a direct physical observable,
    is determined on the lattice, one may simply use the Schr\"{o}dinger
   equation in the infinite space to calculate observables
   such as the scattering phase shifts, bound state spectra etc.  
  Moreover, the potential would be a smooth
   function of the quark masses: This is in sharp contrast to the 
   scattering length which shows a singular 
  behavior around the quark mass corresponding to the 
  formation of the two-body bound states such as the deuteron. 
  Similar situation is well-known in the  
  BEC-BCS crossover of cold  fermionic atoms \cite{BEC-BCS}.
   
  In this article, we will show some recent results of the 
  nuclear force on the lattice (or the lattice nuclear force in short)
  after a brief introduction to our basic formulation.

\section{Deriving the $NN$ potential on the lattice}
\label{sec:strategy}

\subsection{NBS wave function on the lattice}

 In field theory, the best analogue of the two-particle wave function 
 is the equal-time Nambu-Bethe-Salpeter (NBS) amplitude or the``NBS wave function":
  Let us consider an exact  six-quark state $\vert {\cal E} \rangle $ which has
   total energy ${\cal E}$, total three-momentum zero and total electric charge 
   $+e$  in a finite box. Then we define the NBS wave function by
\begin{eqnarray}
\psi(\br)  = 
\langle 0 \vert  {n}_{\beta}(\bx+\br,t=0) {p}_{\alpha}(\bx,t=0) 
\vert {\cal E} \rangle .
\label{eq:BS-def-def}
\end{eqnarray}   
The   local composite operators for the proton and the neutron
 are denoted by   $p_{\alpha}(\bx,t)$ and  $n_{\beta}(\by,t)$
 with spinor indices $\alpha$ and $\beta$.
  One should keep in mind  that $| {\cal E}  \rangle $ is {\it not} 
 a simple superposition of a product state $| {\rm p} \rangle \otimes
|{\rm n} \rangle$,
 since there are complicated exchanges of quarks and gluons between the two composite
 particles.
  The NBS wave function $\psi(\br)$ 
 can be regarded as a probability amplitude in 
  $| {\cal E} \rangle $  to find 
 ``neutron-like'' three-quarks located at point $\bx+\br$ and
 ``proton-like'' three-quarks located at point $\bx$.

 The spatial extent of the $NN$ interaction in QCD is short ranged and is 
 exponentially suppressed beyond the distance $R \sim 2 $ fm.
 Therefore, the spatial part of the NBS wave function in the 
  ``outer region" 
  satisfies the Helmholtz equation, 
 \begin{eqnarray} 
   ( \nabla^2 + k^2 )\psi(\br) =0 \ \ \ \ (|\br| > R),
 \end{eqnarray}
 up to an exponentially small correction in $L$.
 Here the ``asymptotic momentum" $k$ is determined by the 
 asymptotic behavior of the wave function in the outer region.

 An important property of the NBS wave function $\psi(\br)$   
 is that its asymptotic behavior at large $|\br|$ in the infinite volume limit
 reproduces the correct phase shift obtained from 
 the   $S$-matrix of the elastic $NN$ scattering.
 This can be shown explicitly by using the Nishijima-Zimmermann-Haag(NHZ)'s 
  reduction formula \cite{NZH} for the products of local composite operators. 
 (See Appendix A of \cite{Aoki:2009ji} for the proof.)
  To define the NBS wave function on the lattice,
  we start with the four-point function
\begin{eqnarray}
 {\cal G}(\br, t-t_0)= \left\langle 0
   \left|    n_\beta(\bx+\br,t)
    p_\alpha(\bx,t) S(t_0)   \right| 0  \right\rangle 
  \rightarrow   \psi(\br)\  {e}^{-{\cal E}_0(t-t_0)} \ \ (t \gg t_0),
\end{eqnarray}  
 where ${\cal E}_0$ is the lowest energy state  
  created by the source operator $S(t_0)$.

 \subsection{Non-local potential and the velocity expansion}

 To define the $NN$ potential from the 
 NBS wave function, let us introduce the following local function:
 \beq
 K_{E}(\br)= \frac{1}{2\mu} (\nabla^2 + k^2) \psi_{_E}(\br)
 \equiv (E-H_0) \psi_{_E}(\br) .
 \label{eq:kernel}
\eeq 
  In the second equality,  we introduce  an ``effective center of mass  energy",
 $E=k^2/(2\mu)$,  and the free Hamiltonian $H_0=-\nabla^2/(2\mu)$, with
  $\mu=m_N/2$ being the reduced mass of the two nucleons. 
 They are introduced only to make a formal resemblance with the 
 Schr\"{o}dinger type equation and have nothing to do with 
  non-relativistic approximation.  Hereafter, we put the suffix $E$ to
   the NBS wave function to emphasize its $E$-dependence.
 Since the ``plane-wave" part of the NBS wave function in the outer region ($r > R$)
  is projected away by the operator $E-H_0$, 
 the function  $K_{E}(\br)$ is non-vanishing only in the inner region ($r < R$).
  Note also that the Fourier transform of  $K_{E}(\br)$
   is essentially the  half off-shell $T$-matrix.

 We can rewrite Eq.(\ref{eq:kernel})  in two equivalent ways:
 \beq
 (E-H_0) \psi_{_E}(\br) = U_E(\br) \psi_{_E}(\br) = \int U(\br, \br') \psi_{_E}(\br') d\br' .
 \label{eq:potential-2}
\eeq 
The first equality is just a definition of the energy-dependent local
 potential, $U_E(\br)=K_{E}(\br)/\psi_{_E}(\br)$.
  On the other hand,  the energy-independent
  non-local potential,  $U(\br, \br')$,
  is defined  from $U_E(\br)$  through a self-consistent equation,
 \beq
 U (\br,\br')= \langle \br | \hat{U} |\br' \rangle
 =  \sum_E \int_{-\infty}^{+\infty} \frac{dt}{2\pi} 
 U_E(\br) \langle \br | e^{i(\hat{H}_0+\hat{U}-E)t} |\br' \rangle .
 \label{eq:U-U_E}
\eeq
 Carrying out the $t$ integration formally, one may also write
  Eq.(\ref{eq:U-U_E}) as  $\hat{U}= \sum_E \hat{U}_E \delta (E- \hat{H}_0 - \hat{U})$.
 In these formulas, $\sum_E$ stands for the summation (integration) over the 
  discrete (continuum) energies.  In particular, $E$ is always discrete
  on the lattice with a finite volume.  Also, $E$ has an upper limit
   $E_c$ at which inelastic scattering starts to take place.
      Eliminating the 
  $E$-dependence of the potential 
   through Eq.(\ref{eq:U-U_E}) has been discussed in a transparent manner by 
   Kr\'olikowski and  Rzewuski \cite{KR56} long time ago: their motivation was to  
    prove the equivalence between the multiple-time
     Nambu-Bethe-Salpeter type equation with an $E$-dependent kernel
     and the equal-time Schr\"{o}dinger type equation with an $E$-independent 
   potential. Essentially the same method was
   rediscovered and discussed  in \cite{Ishii:2006ec,Aoki:2008hh,Aoki:2009ji}
    in the context of the  NBS wave function on the lattice.   
 
 If we further focus on the 
  low-energy scattering with $E$ sufficiently smaller than
   the intrinsic scale of the system or the 
    scale of the non-locality of the potential in  Eq.(\ref{eq:potential-2}),
   the velocity expansion
   of $U(\br,\br')$ in terms of its  non-locality is useful \cite{TW67}:
  For example, the potential with hermiticity, 
   rotational invariance, parity symmetry,
   and time-reversal invariance may be expanded as \cite{okubo}
 \beq
\label{eq:U-del}
  U(\br,\br')    &=& V(\br, \bv) \delta(\br-\br'), \\
 V(\br, \bv)    & =&
   \underbrace{V_C(r) + V_T(r) S_{12}}_{\rm LO} 
  + \underbrace{V_{LS}(r) {\bL} \cdot {\bS}}_{\rm NLO}  
  +\underbrace{{O}(\bv^2)}_{{\rm N}^2{\rm LO}}
  + \cdots , 
   \label{eq:V-pot}
 \eeq  
   where $\bv = \bp/\mu $ and $\bL = \br \times \bp $ with 
   $\bp = -i \nabla$, and 
   $S_{12}=3(\bsigma_1 \cdot \br)(\bsigma_2 \cdot \br)/r^2 - \bsigma_1 \cdot \bsigma_2$.
  Each coefficient of the expansion is  
 a local potential and  can be determined  successively 
 by measuring the NBS  wave functions for several different energies.
 The central potential $V_C$ and 
  the tensor potential $V_T$ are classified as
  the leading order (LO) potentials since they
  are of $O(\bv^0)$. The next-to-leading (NLO) potential of  
  $O(\bv)$ is the spin-orbit potential  $V_{LS}(r)$. 
  The LO and NLO potentials are phenomenologically known to be the 
   dominant  interactions
    at low energies.
   
 An advantage of defining the potential 
 from the NBS wave functions in the ``inner region" is that
 the effect of the lattice boundary is 
 exponentially suppressed for finite range interactions:
  Then one can first make 
  appropriate extrapolation of
     $U(\br,\br')$ or $V(\br, \bv)$  to $L\rightarrow \infty$, and then
 solve the Schr\"{o}dinger equation  using the extrapolated potential
   to calculate the observables such as 
    the phase shifts and binding energies in the 
    infinite volume. 
   This is in contrast to the  L\"{u}scher's approach \cite{luescher} in
   which  the wave functions in the ``outer region" suffering from 
   the boundary conditions is ingeniously utilized to probe the 
  scattering observables.  Apparently, 
  the two approaches are  the opposite
  sides of a same coin.

\subsection{Interpolating operator and the potential}

In Eq.~(\ref{eq:BS-def-def}),
 simplest interpolating operators  for the neutron and 
the proton written in terms of the up-quark 
 $u(x)$ and the down-quark  $d(x)$ would be 
\begin{eqnarray} 
   n_\beta(x) =
  \varepsilon_{abc} \left(
		     u_a(x) C \gamma_5 d_b(x)
		    \right)
  d_{c\beta}(x), \ \ 
   p_\alpha(x) =
  \varepsilon_{abc} \left(
		     u_a(x) C \gamma_5 d_b(x)
		    \right)
  u_{c\alpha}(x),
\end{eqnarray}   
where $x=(\bx, t)$ and  
 the color indices are denoted by $a$, $b$ and $c$. The
 charge conjugation matrix in the spinor space is denoted
 by $C$.
 The local operators given above are
   most convenient for relating  the NBS wave function to the 
   four-point Green's function and the scattering 
   observables at $L \rightarrow \infty$ through the NZH reduction formula. 

  In principle,  one may choose any composite operators
  with the same quantum numbers as the nucleon to define the NBS wave
  function.  Different interpolating operators lead to different
  NBS wave functions and different $NN$ potentials. However, they lead to
  the same physical observables  by construction.
   Analogous situation can be seen in quantum mechanics where the unitary
   transformations  modify both the wave function and the potential 
   in such a way that observables are unchanged.   
   Even more direct analogy is in field theory for point-like particles:
   Field re-definitions 
   modify the vertices and propagators in the Feynmann rule, while
   the on-shell $S$-matrix is not affected by such changes.

\subsection{Central  and tensor forces}  
\label{sec:tensor}

 In the LO of the  velocity expansion in Eq.~(\ref{eq:V-pot}),
 we have the central potential $V_C(r)$  and  the tensor potential  
$V_T(r)$, so that the Shr\"{o}dinger equation reads
\beq
  \bigl( E- H_0 \bigr)
  \psi_{_E}(\br)
  =
   \bigl( V_{C}(r) + V_{T}(r) S_{12} \bigr)
  \psi_{_E}(\br).
  \label{schrodinger.eq.one.plus}
\eeq 
The central potential  acts
separately on the orbital S-state and the D-state, while  the  tensor potential
 provides a coupling between these two. Therefore,
  a coupled-channel Schr\"odinger equation  
  is obtained from Eq.(\ref{schrodinger.eq.one.plus})
 by using the projection operators ${\cal P}$ and ${\cal Q}$
 to the S-state and D-state, respectively. 
 Eventually we calculate $V_C$ and $V_T$ from the following formula
 where the quantities in the right hand side are all known on the lattice:
\beq
\left(
  \begin{array}{c}
      V_C     \\
      V_T     \\
  \end{array}
\right)
= \left(
  \begin{array}{cc}
    {\cal P}\psi_{_E}    & {\cal P}S_{12} \psi_{_E}    \\
    {\cal Q} \psi_{_E}   & {\cal Q}S_{12} \psi_{_E}  \\
  \end{array}
\right)^{-1} 
\left(
  \begin{array}{cc}
   E-H_0    &  0       \\
     0      &  E-H_0   \\
  \end{array}
\right)
\left(
  \begin{array}{c}
      {\cal P}\psi_{_E}     \\
      {\cal Q}\psi_{_E}     \\
  \end{array}
\right).  
\eeq

\section{Numerical results in quenched and full QCD simulations}
\label{sec:numerical} 

\subsection{LO potentials}
\label{sec:LO-potentials}

\begin{figure}[t]
\begin{center}
\includegraphics[width=5.5cm,angle=-90]{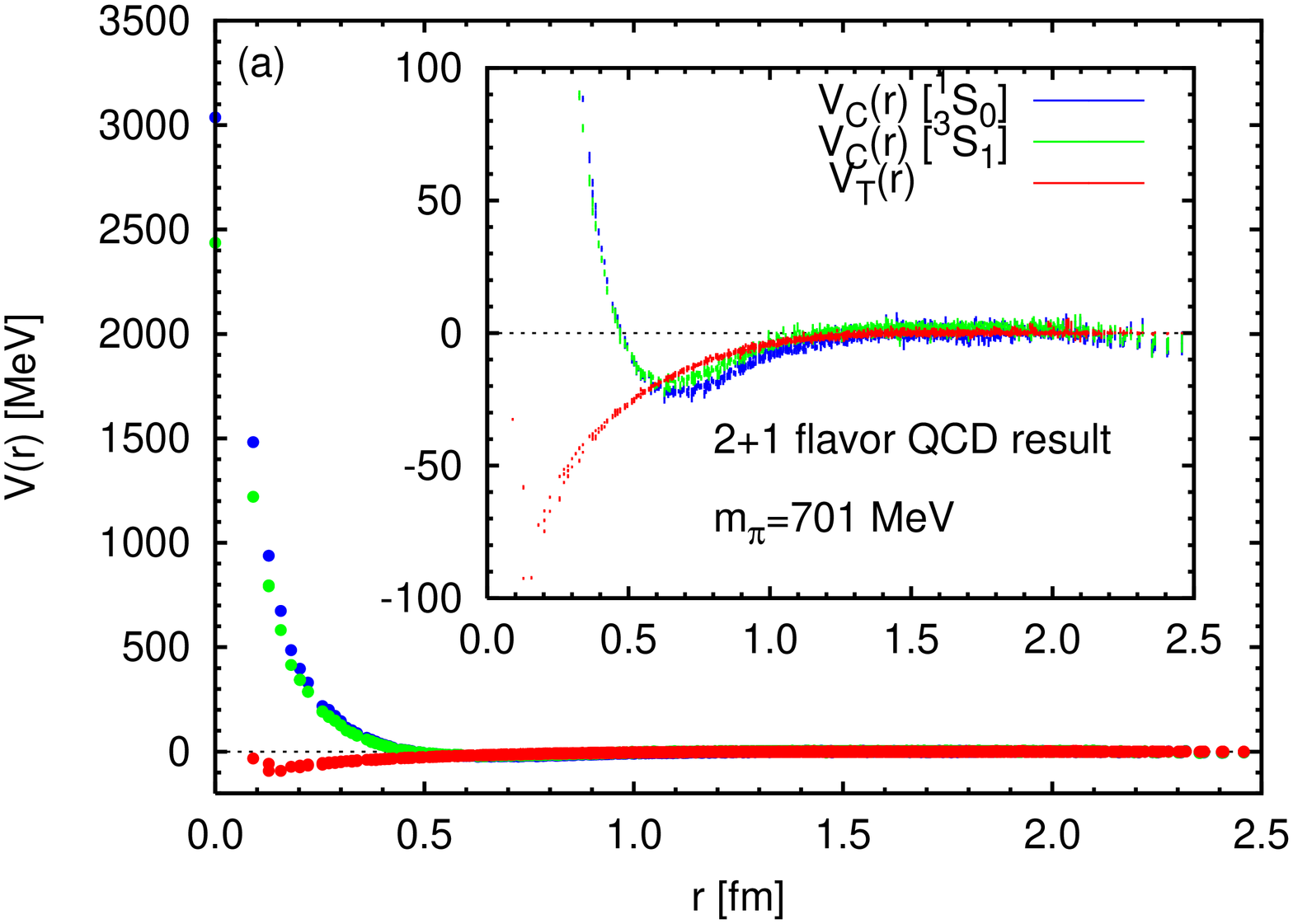}\hfill
\includegraphics[width=5.5cm,angle=-90]{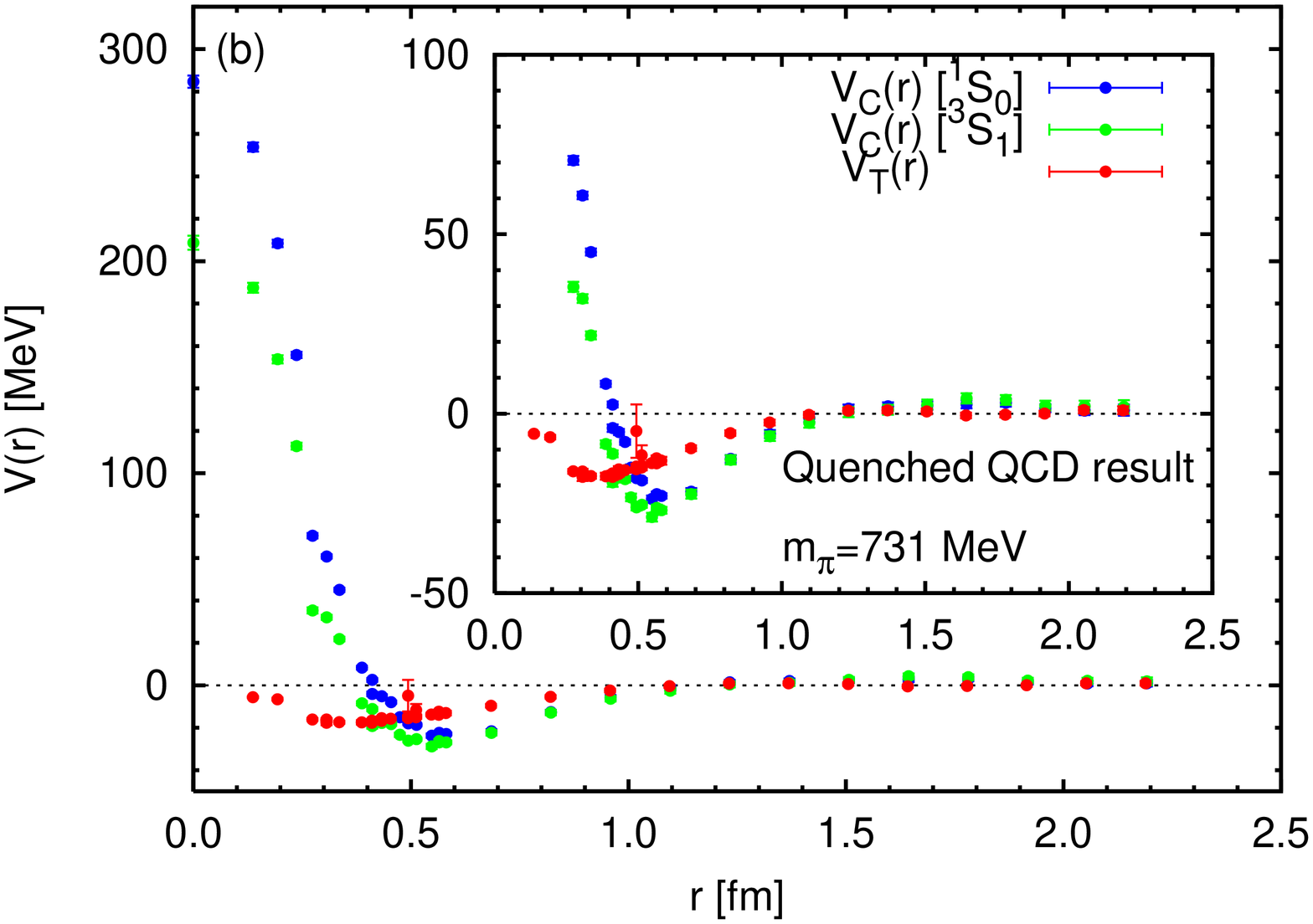}
\end{center}
\caption{(a) LO potentials in (2+1)-flavor QCD for $m_{\pi}$=701 MeV \cite{Ishii:2010th}.
 (b) LO potentials in quenched QCD for $m_{\pi}$=731 MeV \cite{Aoki:2009ji}. }
\label{fig:LO-pot}
\end{figure}

\begin{figure}[t]
\begin{center}
\includegraphics[width=3.7cm,angle=-90]{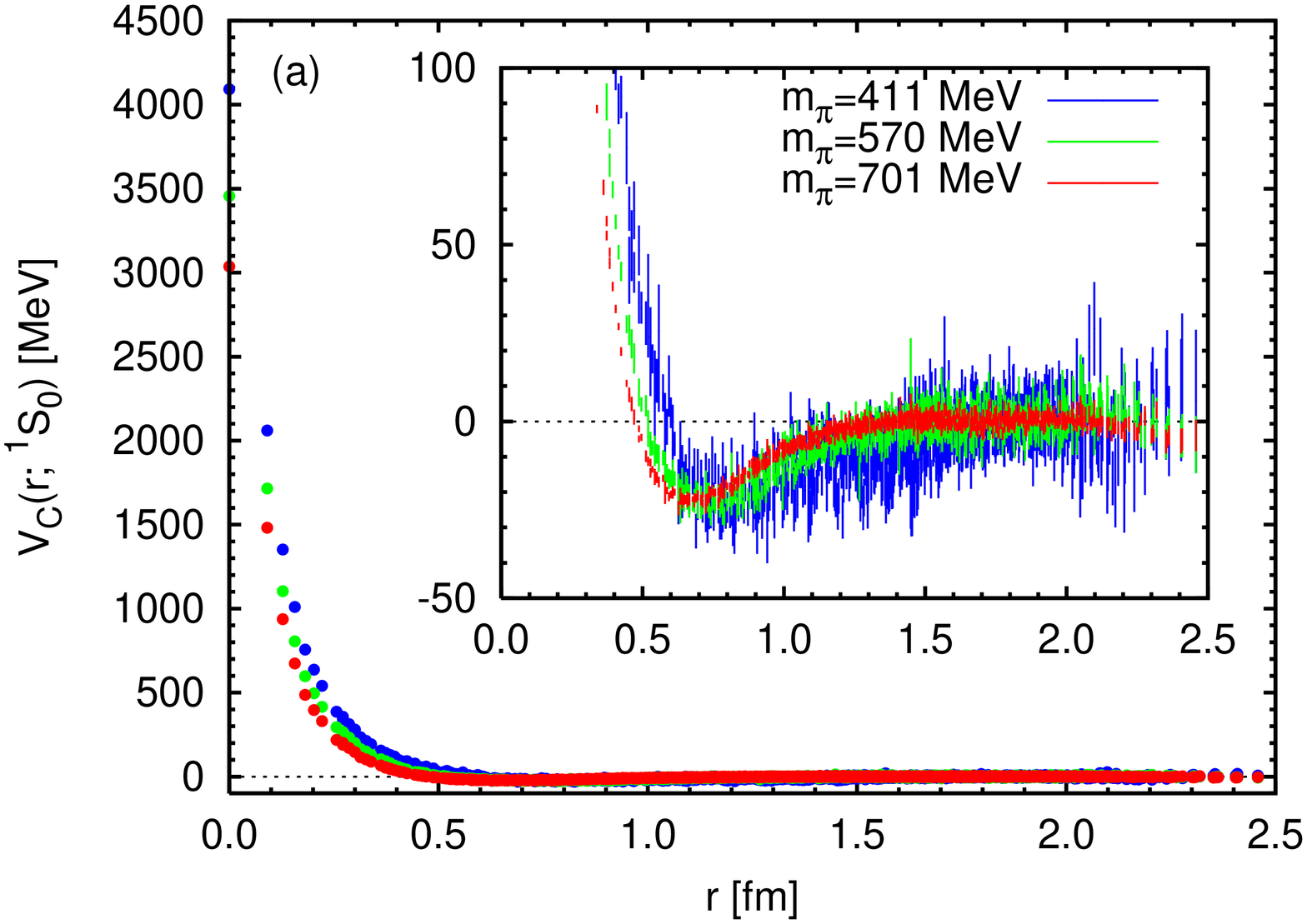}\hfill
\includegraphics[width=3.7cm,angle=-90]{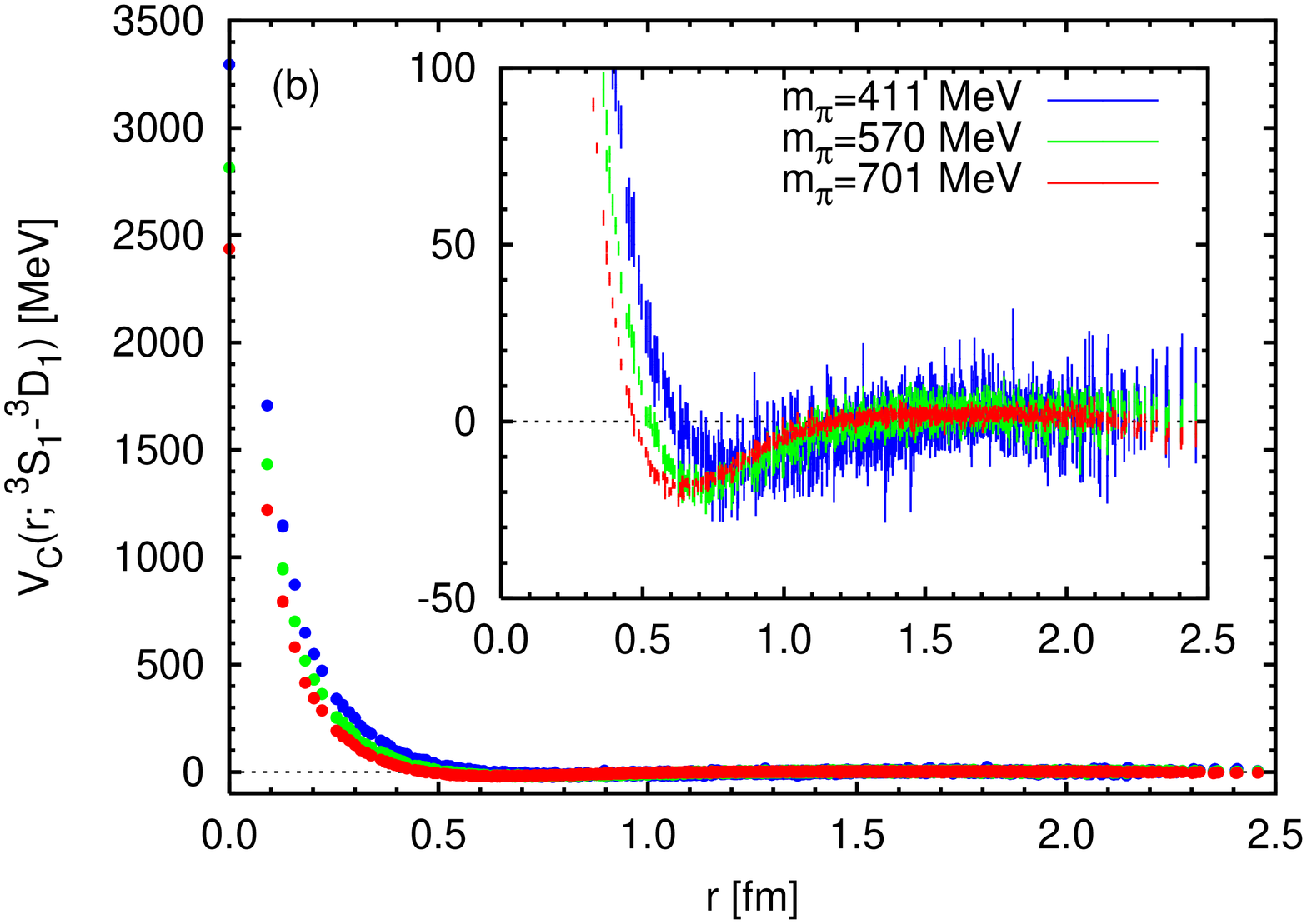}\hfill
\includegraphics[width=3.7cm,angle=-90]{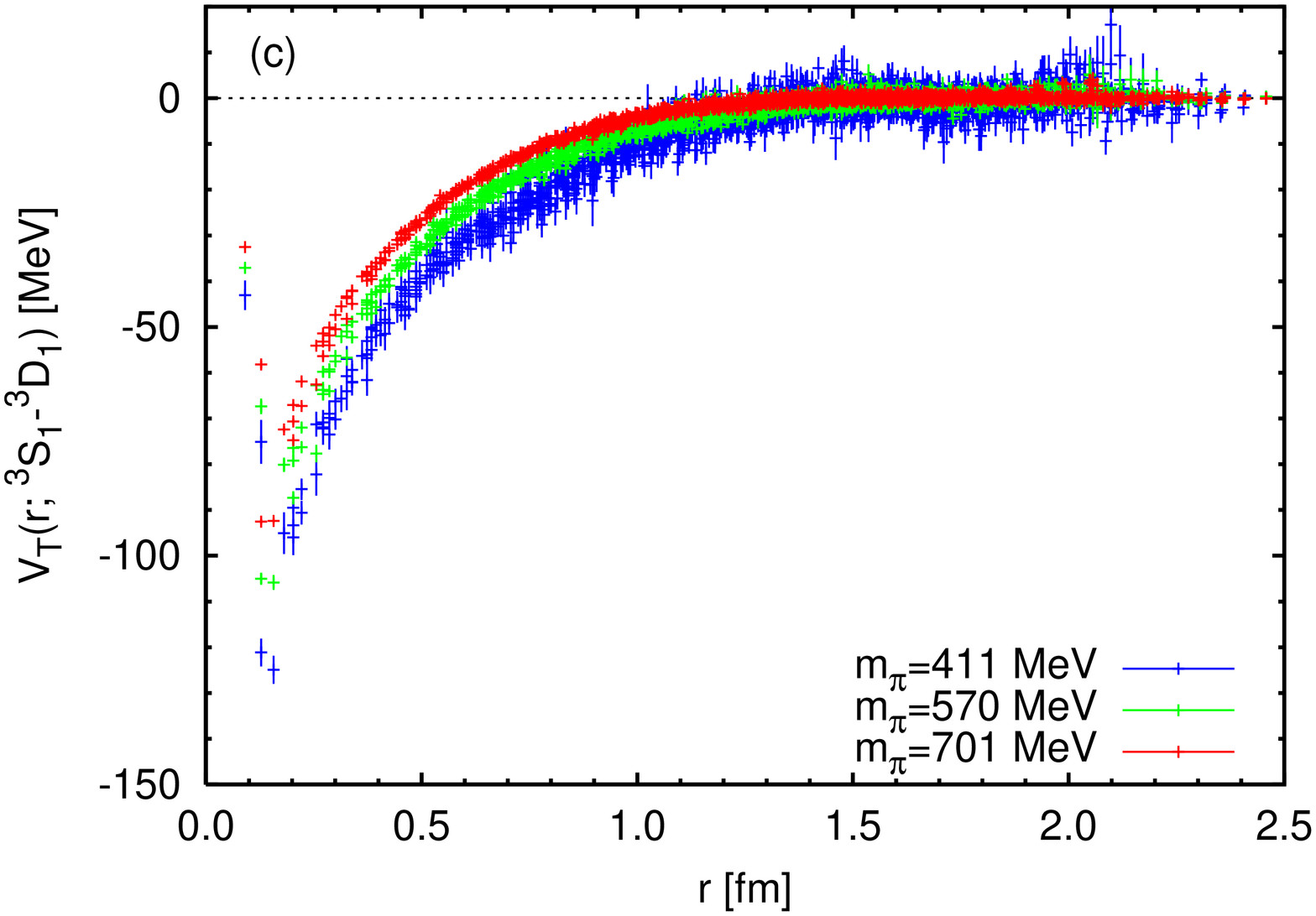}
\end{center}
\caption{Quark mass dependence of the LO potentials in (2+1)-flavor QCD.
(a) The central potential in the spin-singlet channel, 
(b) the central potential in the spin-triplet channel, and 
(c)  the tensor potential in the spin-triplet channel \cite{Ishii:2010th}.}
\label{fig:quark-pot}
\end{figure}

 To demonstrate whether the formalism discussed in the previous section indeed works,
 we first carried out a quenched QCD simulations with the standard plaquette gauge
action and the standard Wilson quark action
 on a $32^3\times 48$ lattice \cite{Ishii:2006ec,Aoki:2008hh,Aoki:2009ji}.
  The lattice spacing is $a =0.137$ fm 
which corresponds to the spatial size  $L= 4.4$ fm. The light 
quark masses are chosen so that 
 we have $m_{\pi}=$731, 529 and 380 MeV and $m_{N}=$1558, 1334 and 1197 MeV,
respectively.   Periodic or anti-periodic boundary conditions are imposed
on the quark field along the spatial direction.

 As for full QCD with the vacuum polarization of quarks included,
 we use the PACS-CS gauge configurations in (2+1)-flavor QCD
  generated by the Iwasaki gauge action 
 and the $O(a)$-improved Wilson quark (clover) action
on a $32^3\times 64$ lattice \cite{Aoki:2008sm}. 
The lattice spacing is $a =0.091$ fm 
which corresponds to the spatial size  $L= 2.9$ fm.
The light quark masses are chosen so that 
 we have $m_{\pi}=$701, 570 and 411 MeV and $m_{N}=$1583, 1412 and 1215 MeV,
respectively.   Also, Periodic  boundary condition is imposed
on the quark field along the spatial direction.

 Shown in Fig.\ref{fig:LO-pot}(a) are the LO potentials 
 ($V_C$ for $^1S_0$ and $^3S_1$ channels and $V_T$ determined from
 $^3S_1$-$^3D_1$ channel) in (2+1)-flavor QCD
 for $m_{\pi}$=701 MeV. Even with such a  large quark mass,  
 there is a clear evidence of the 
  repulsive core surrounded by attractive well for for central potential
   and an evidence of a mild  tensor force  \cite{Ishii:2010th}. They have qualitative
    similarity with phenomenological potentials.
We show in Fig.\ref{fig:LO-pot}(b) the LO potentials in quenched QCD
  with $m_{\pi}$=731 MeV for comparison \cite{Aoki:2009ji}.
  Although the qualitative structure of the potentials are
   the same, the magnitude of the 
   repulsive core and the tensor force are relatively weak in quenched QCD.

Shown in Fig.\ref{fig:quark-pot}(a,b,c) are the quark mass dependence
of the LO potentials in (2+1)-flavor QCD \cite{Ishii:2010th}.
 As the quark mass decreases,
 the repulsive core in (a,b) and the tensor force in (c)
  become stronger and the  attractive well in (a,b) becomes wider.  
 We have fitted these potentials and have calculated the 
  $NN$ scattering phase shift by solving the Schr\"{o}dinger equation.
   We found that deuteron bound state does not appear for
    these quark masses, so that further reduction of the quark mass
    would be necessary to obtain the realistic lattice potentials.

\subsection{Convergence of the velocity expansion}
\label{sec:NNLO-pot}

\begin{figure}[t]
\begin{center}
\includegraphics[width=0.49\textwidth]{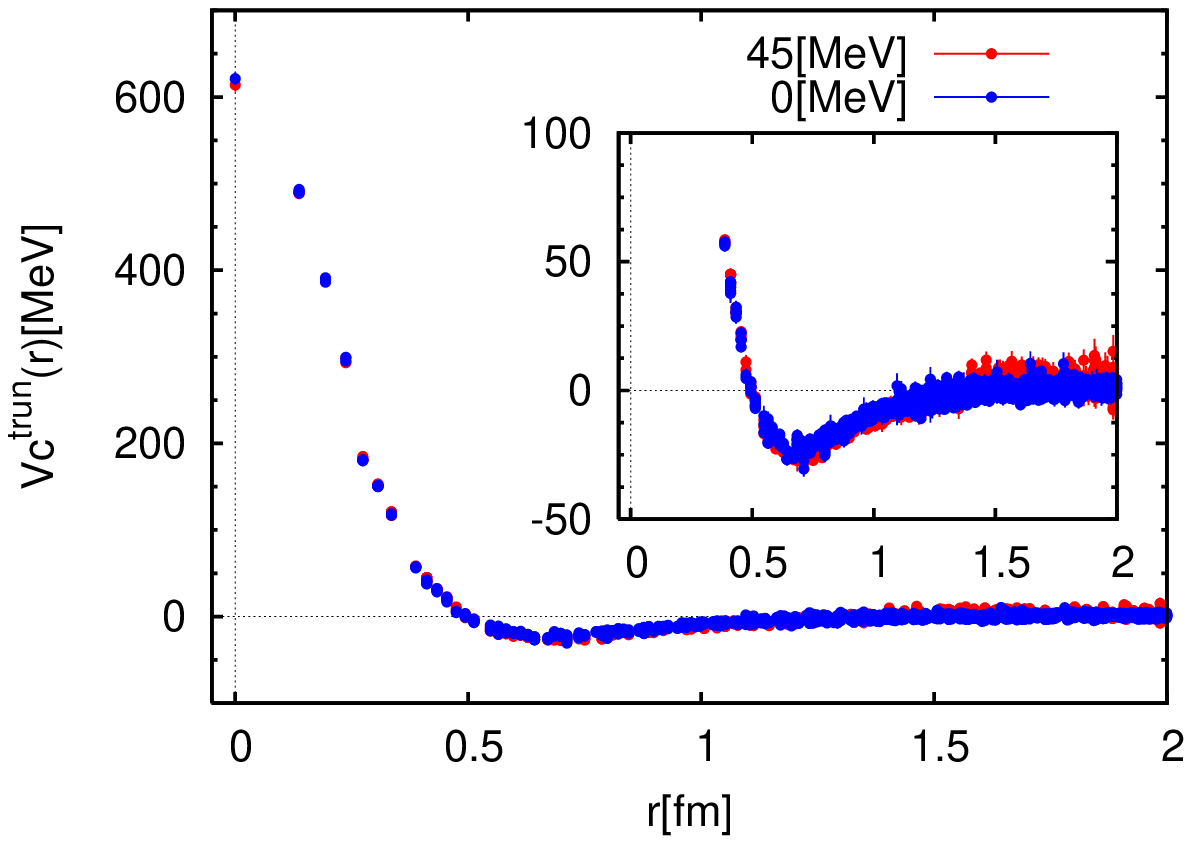}
\includegraphics[width=0.49\textwidth]{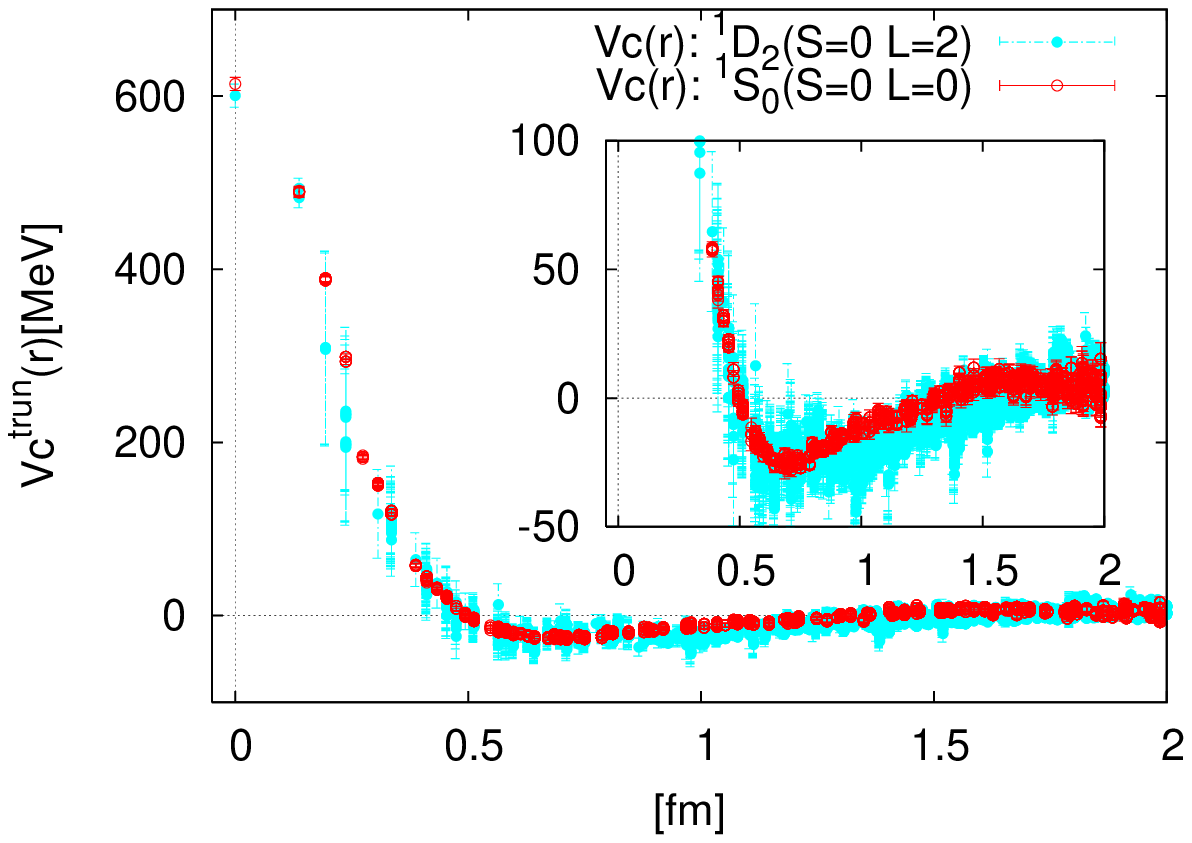}
\end{center}
\caption{(Left) A comparison of the $^1S_0$ central potentials obtained 
at different energies ($E\sim 0$ MeV vs. $E\sim 45$ MeV).
 (Right) A comparison of the spin-singlet central
potential at $E \sim$ 0 MeV with different orbital angular momenta ($L=0$ vs. $L=2$). 
 Figures are taken from \cite{Murano:2011}.}
\label{fig:quark-pot2}
\end{figure}

  So  far, the potentials are derived with the periodic boundary condition
  in the spatial direction for the quark fields.  This leads to the 
  the ``effective center of mass  energy" $E=k^2/(2\mu)$ almost zero. 
  To study the convergence of the velocity expansion of the 
  non-local potential in Eq.(\ref{eq:V-pot}), we compare the 
  local $^1S_0$ potential (in quenched QCD  with 
  $m_{\pi}=529$ MeV) obtained at $E\simeq 0$ MeV under
  the periodic boundary condition and that obtained at 
  $E\simeq 45$ MeV under  the anti-periodic boundary condition \cite{Murano:2011}.
  Good agreement between the two as shown in 
  the left panel of Fig.\ref{fig:quark-pot2} indicates that a $O(\bv^2)$ term
  in the  N$^2$LO level is rather small in this energy interval.
  Shown in the right panel of  Fig.\ref{fig:quark-pot2} is a test for a different
  $O(\bv^2)$ term in the  N$^2$LO level \cite{Murano:2011}: In this case, 
   local potentials determined at the same energy ($E \simeq 0$ MeV) 
     with different orbital angular momenta ($L=0,2$)
    in the spin-singlet channel are compared.  Again, within statistical
    errors, the effect of the N$^2$LO term is likely to be small.
 
\section{Hyperon interactions}
\label{sec:hyperon}

To unravel the origin of the repulsive core in the $NN$ interaction,
let us consider the $S$-wave interaction between octet baryons  
 in the flavor SU(3) limit. In this case, 
two baryon states with a given angular momentum
are labeled by the irreducible flavor multiplets as
\begin{equation}
 {\bf 8} \otimes {\bf 8} 
 = \underbrace{{\bf 27} \oplus {\bf 8}_s \oplus {\bf 1}}_{\mbox{symmetric}} ~ 
  \oplus \underbrace{{\bf 10}^* \oplus {\bf 10} \oplus {\bf 8}_a}_{\mbox{anti-symmetric}} \ . 
\end{equation}
Here ``symmetric" and ``anti-symmetric" stand for the symmetry under the
flavor exchange of two baryons.
For the system in the orbital S-wave, the Pauli principle between two baryons imposes 
${\bf 27}$, ${\bf 8}_s$ and ${\bf 1}$ to be spin singlet  ($^1S_0$) while 
${\bf 10}^*$, ${\bf 10}$ and ${\bf 8}_a$ to be spin triplet ($^3S_1$). 
Since there are no mixings among different multiplets in the SU(3) limit, 
one can define the corresponding potentials as
 \begin{eqnarray}
^1S_0 \ &:& \  V^{({\bf 27})}(r), \ V^{({\bf 8}_s)}(r), \ V^{({\bf 1})}(r), 
\\ 
^3S_1 \ &:& \ V^{({\bf 10}^*)}(r), \ V^{({\bf 10})}(r), \ V^{({\bf 8}_a)}(r) ~.
\end{eqnarray}
Potentials among octet baryons, both the diagonal part ($B_1B_2 \rightarrow B_1 B_2)$ and  
the off-diagonal part ($B_1B_2 \rightarrow B_3 B_4$), are obtained by  suitable combinations
of $V^{(\alpha)}(r)$ with $\alpha={\bf 27},{\bf 8}_s,{\bf 1},{\bf10}^*,{\bf 10},{\bf 8}_a$.

In this SU(3) study, we employ the gauge configurations on a $16^3 \times 32$ lattice
generated by CP-PACS and JLQCD Collaborations with the renormalization group improved
 Iwasaki gauge action and the non-perturbatively $O(a)$ improved Wilson quark 
 action. The lattice spacing and the lattice volume are
 $a=0.121(2)$ fm and $L=1.93(3)$ fm, respectively.
These configurations are provided by Japan Lattice Data Grid (JLDG) 
and International Lattice Data Grid (ILDG) \cite{CPPACS-JLQCD}.

\begin{figure}[tp]
\begin{center}
 \includegraphics[width=0.4\textwidth]{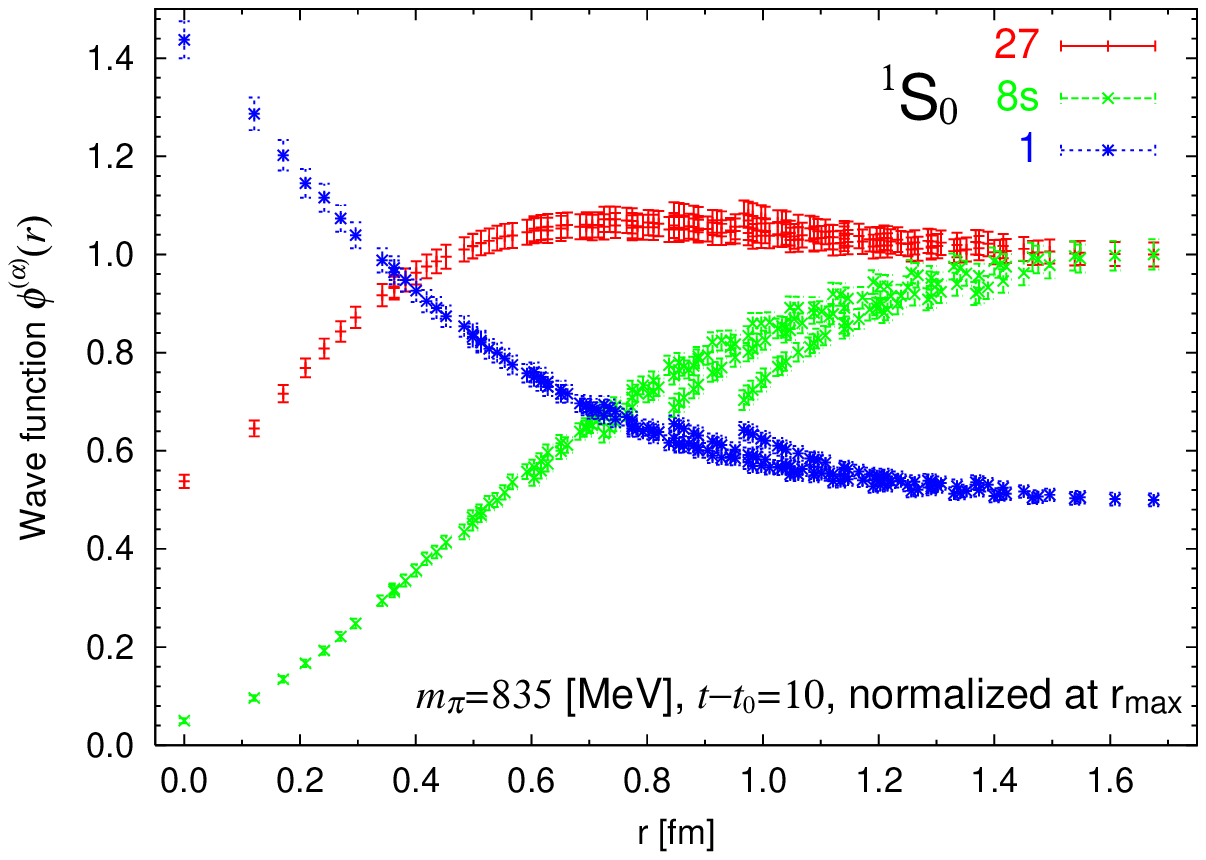} \ \ \ \ \ 
 \includegraphics[width=0.4\textwidth]{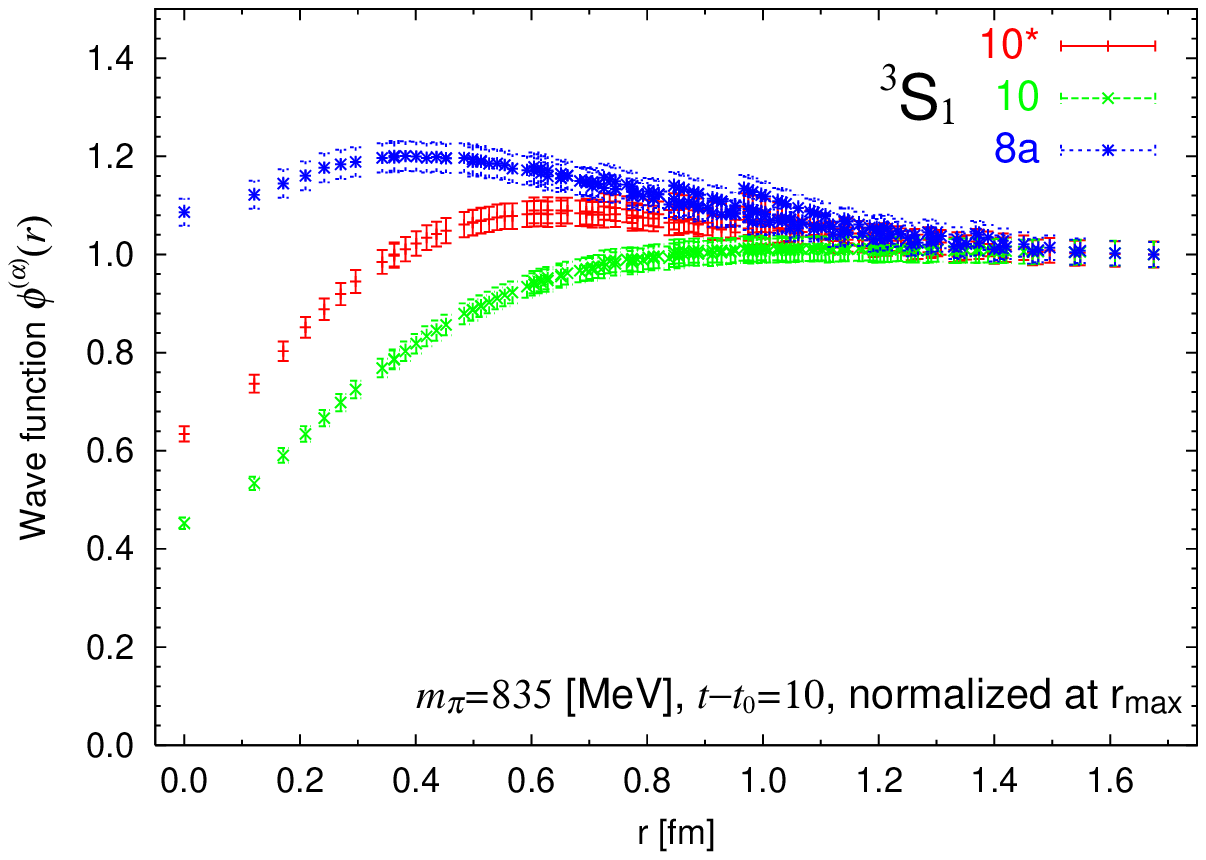}
\end{center}
\caption{\label{fig:wave}
 NBS wave functions  at $m_{\pi}=835$ MeV, normalized to 1/2 for the singlet channel
 and to 1 for other channels at the maximum distance \cite{Inoue:2010hs}.}
\end{figure}

\begin{figure}[tp]
 \includegraphics[width=0.33\textwidth]{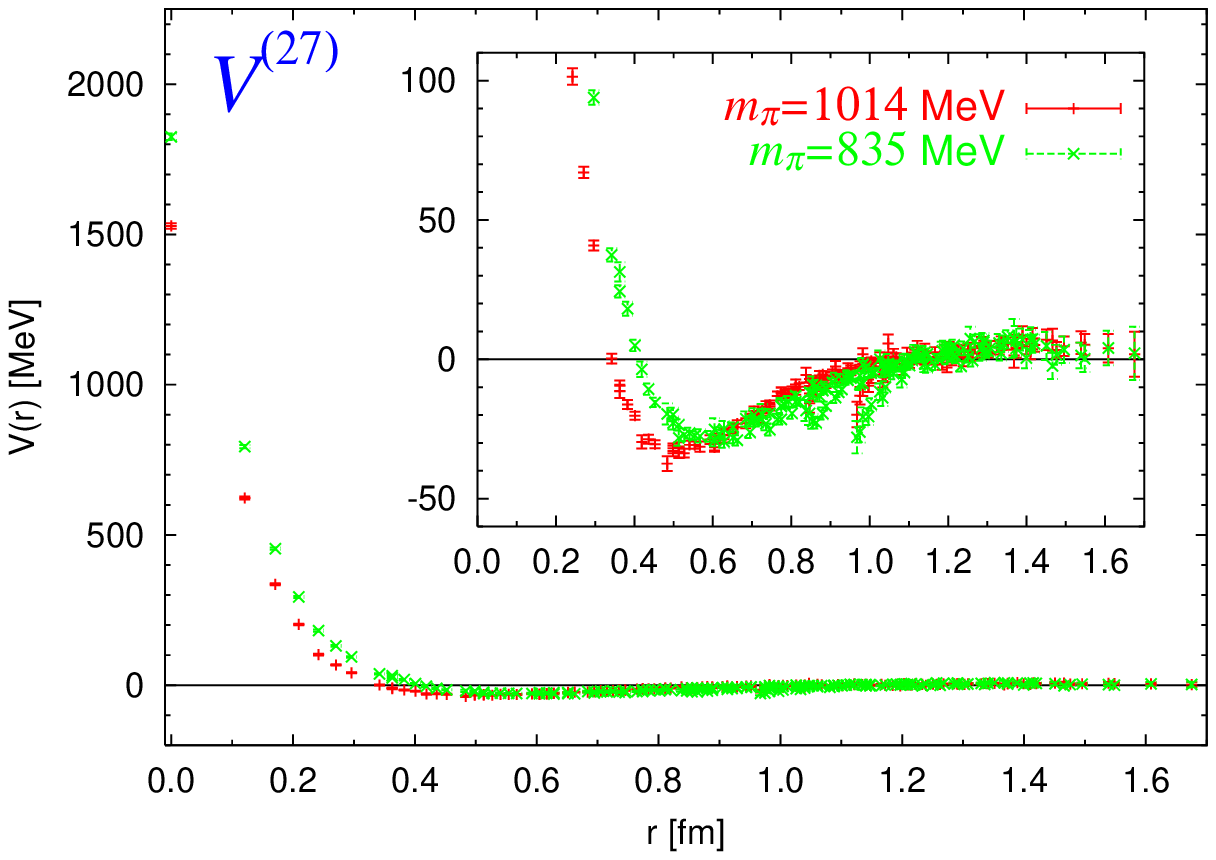}\hfill
 \includegraphics[width=0.33\textwidth]{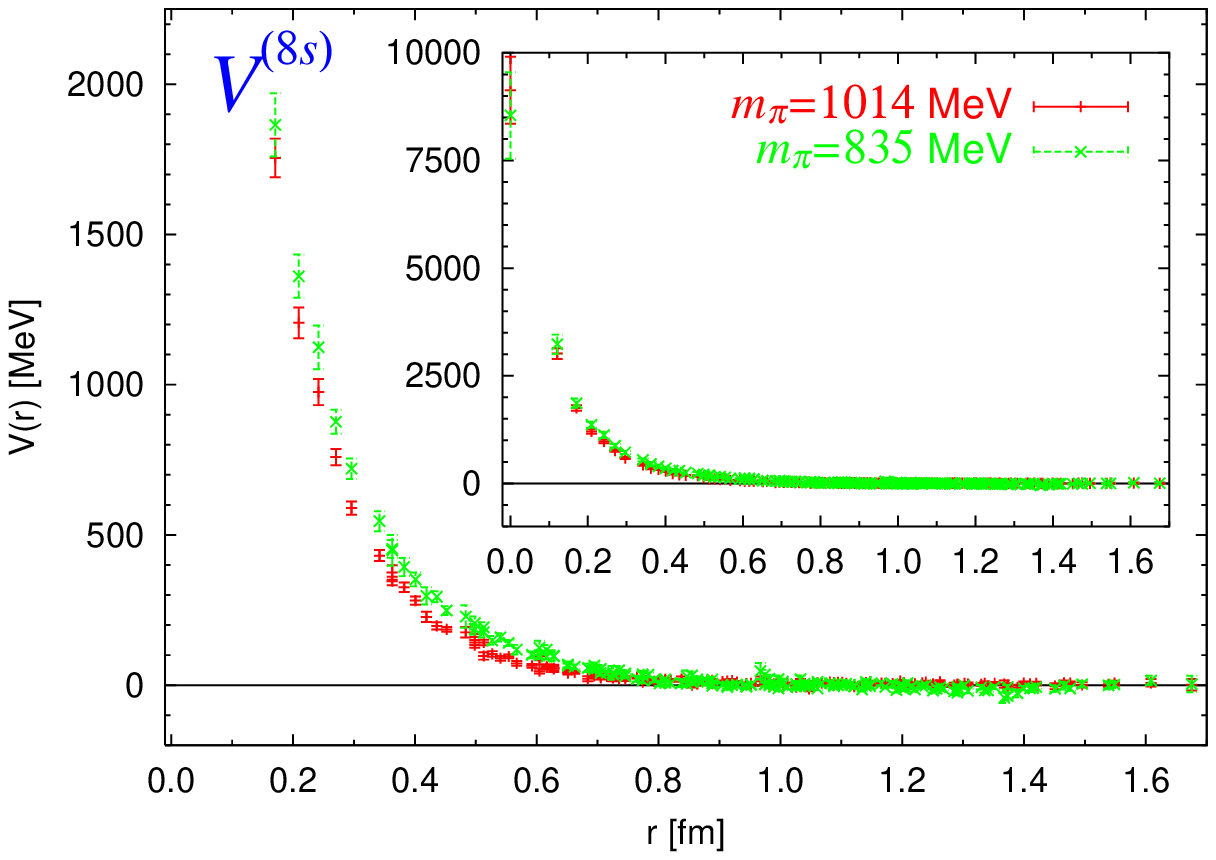}\hfill
 \includegraphics[width=0.33\textwidth]{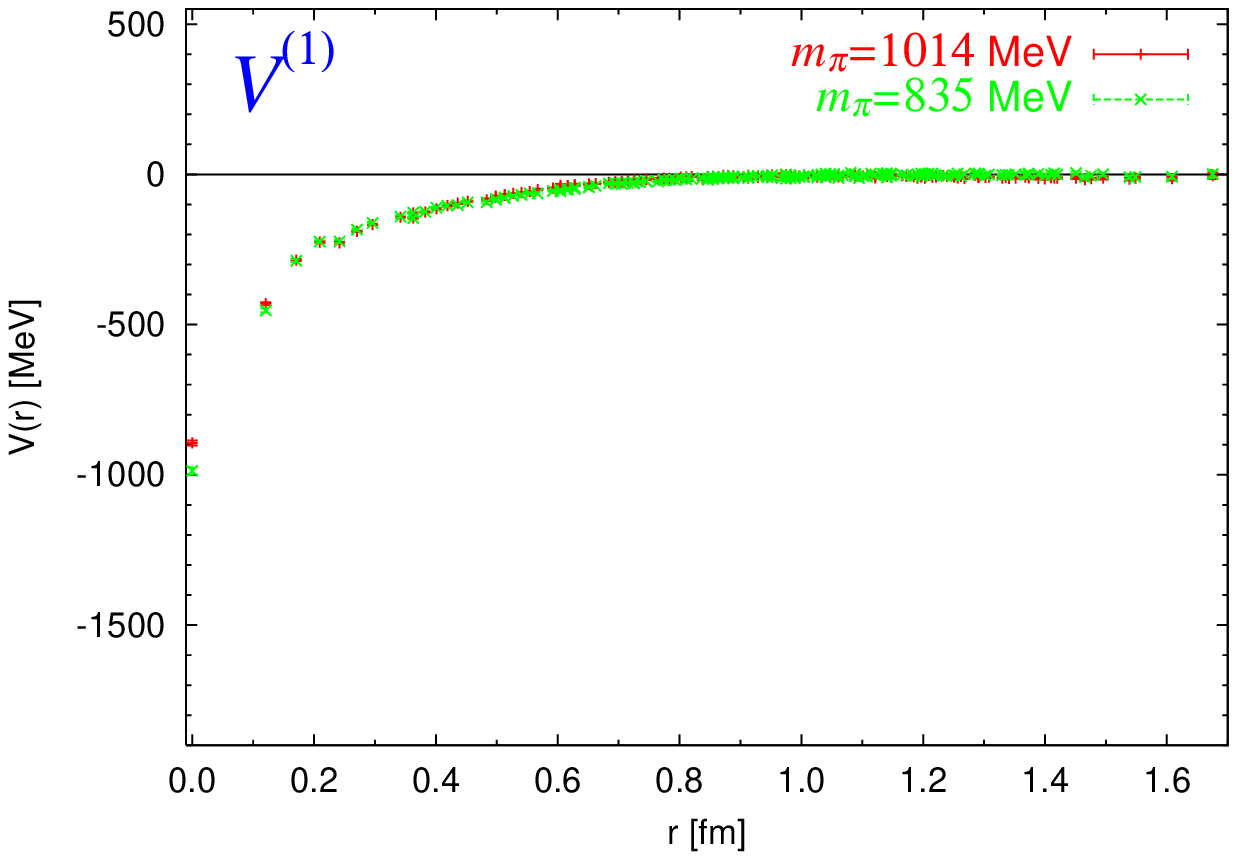}\hfill
\caption{\label{fig:pot_2kappa}
  The three independent BB potentials in the $^1S_0$ channel
   in the flavor SU(3) limit, 
  extracted from the lattice QCD simulation at 
  $m_{\pi}=1014$ MeV (red bars) and $m_{\pi}=835$ MeV (green crosses).
 } 
\end{figure}

Fig. \ref{fig:wave} shows the NBS wave functions as a function
of the relative distance between two baryons 
at  $m_{\pi}=835$ MeV \cite{Inoue:2010hs}. 
To draw all data in a same scale, they are normalized to 1/2 for the singlet channel
and to 1 for other channels at the maximum distance. 
The wave functions in Fig. \ref{fig:wave} show characteristic flavor dependence:
In particular, a strong suppression at short distance appears in the ${\bf 8}_s$ channel,
while a strong enhancement appears in the ${\bf 1}$ channel.
Similar results are obtained for $m_{\pi}=1014$ MeV too.

Fig. \ref{fig:pot_2kappa} shows the resulting three independent BB potentials
in the $^1S_0$ channel in the flavor basis obtained from the NBS wave functions.
Red bars (green crosses) data correspond to the pion mass 1014 MeV (835 MeV):
Although there is a tendency that the magnitude (range) of the  potentials becomes
larger at short distances (longer at large distances) for lighter quark mass,
the differences are not substantial for the present heavy quark masses. 
  Left panels of Fig. \ref{fig:pot_2kappa} show $V^{({\bf 27})}$ 
  which  corresponds to NN $^1S_0$ potential.
 It has a repulsive core at short distance and an attractive pocket as we have 
  shown already  in quenched and (2+1)-flavor simulations.
The middle panel of Fig. \ref{fig:pot_2kappa} corresponding to
 $V^{({\bf 8}_s)}$   has
a very strong repulsive core among all  channels.
In contrast, the right panel of   Fig. \ref{fig:pot_2kappa} corresponding to
$V^{({\bf 1})}$  shows attraction for all distances, which is relevant to
 the cereblated $H$-dibaryon \cite{Jaffe:1976yi}.

These features are consistent with what has been 
observed in phenomenological quark models \cite{Oka-Fujiwara}.
In particular, the potential in the ${\bf 8}_s$ channel in quark models
becomes strongly repulsive
at short distance since the six quarks cannot occupy the same orbital state
 due to quark Pauli blocking.  On the other hand,
 the potential in the ${\bf 1}$ channel
does not suffer from the quark Pauli blocking and can become attractive
due to short range gluon exchange. 
 Such an agreement between the lattice data and the 
 phenomenological models indicates that the quark Pauli blocking plays an 
 essential role for the repulsive core in BB systems as  
 suggested long time ago in \cite{Otsuki:1965yk}. 
 One can also confirm the idea of the Pauli blocking by considering the 
  meson-baryon interaction such as charmonium-nucleon potential
  \cite{Kawanai:2010ev} and  
  kaon-nucleon potential \cite{Ikeda:2010sg} within the present lattice approach.
  Generalization of the baryon-baryon interaction
  to the case with explicit SU(3) breaking is also under way \cite{Nemura:2008sp}.

\section{Summary and concluding remarks}
\label{sec:summary}

 In this paper, we have discussed  
  the basic notion of the nucleon-nucleon potential and its field-theoretical 
  derivation from the   equal-time Nambu-Bethe-Salpeter wave function in QCD. 
  By construction, the non-local potential defined through the projection of the
   wave function to the interaction region (the inner region)   correctly reproduces the 
   asymptotic form of the wave function in the region beyond the 
   range of the nuclear force (the outer region). Thus the observables such as the 
   phase shifts and the binding energies can be
   calculated after extrapolating the potential to the infinite volume limit.
   Non-locality of the potential can be taken into account successively by
    making its velocity expansion, which introduces the velocity-dependent
     local potentials. The leading-order terms  of such velocity expansion 
      for the 
      nucleon-nucleon interaction are the central and the tensor potentials.
           
  To show how this formulation works, some results in the 
    quenched and (2+1)-flavor lattice  QCD simulations are shown
     for relatively heavy pion masses, $m_{\pi} \sim 400, 500, 700 $ MeV.
   We  found that the $NN$ potential calculated on the lattice at low energy 
  shows all the characteristic features 
  expected from the empirical $NN$ potentials obtained from the experimental
   $NN$ phase shifts, namely  the 
  attractive well  at  long and medium distances and the repulsive core 
  at short distance for the 
   central potential. As for the tensor potential obtained from the  
  coupled channel treatment of the $^3{\rm S}_1$-state
  and the $^3{\rm D}_1$-state,
 we found appreciable attraction at long and medium distances.
 
 As the quark mass decreases, the repulsive core and 
   attractive well in the central potential, and the attractive well in the 
   tensor  potential tend to be enhanced. To make the deuteron bound state, however,
    it is necessary to go the lighter quark masses. 
  We have also shown that the derivative expansion 
   in terms of the local and energy-independent potentials works well 
   at low energies for at least the quark masses studies above.  
   
  There are a number of  directions to be investigated
   on the basis of our approach. Among others, the most 
 important direction  is to carry out 
  (2+1)-flavor simulations with a large volume (e.g. $L=6$ fm) at
   physical quark mass ($m_{\pi}=135$ MeV)  to extract the 
    realistic $NN$ potentials.  This will be indeed started  soon as a first
     priority simulation at 10 PFlops national supercomputer ``KEI"  
   which will have full operation in 2012 at  Advanced Institute for Computational
    Science (AICS) in Kobe, Japan \cite{AICS}.
 Simulations of the three or more nucleons  on the lattice  are also a challenging
  problem to be studied in relation to the attractive binding of finite nuclei
   and to the repulsive effect in high density matter relevant to neutron stars.
  Study along this line has been recently started \cite{Yamazaki:2009ua,Doi:2010yh}.

    If it turns out that the program described in this paper indeed works in lattice QCD
     with the physical quark mass, it would be 
      a major  step toward the understanding 
   of atomic nuclei and neutron stars   from the fundamental law of the strong
   interaction, the quantum chromodynamics.

 \vspace{0.5cm}
 
\noindent
{\bf Note added}: Recently, full QCD simulations of the hyperon potentials
 in the flavor SU(3) limit reported in \cite{Inoue:2010hs} and
 discussed in Sec. \ref{sec:hyperon} 
 were extended to the lattice sizes $L\simeq 3$ and 4 fm
  for the pseudo-scalar meson mass of 673--1015 MeV. 
   By solving the 
  Schr\"{o}dinger eqaution with the flavor-singlet potential,
   a bound $H$-dibaryon with the 
 binding energy of 30--40 MeV was found
   \cite{Inoue:2010es}. Since the binding energy turns out to be
    insensitive to the quark masses, there may be a possibility of weakly bound or
resonant $H$-dibaryon even in the real world with lighter
quark masses and with the flavor SU(3) breaking. To make
a definite conclusion, however, the (2+1)-flavor lattice
QCD simulations for $H$-dibaryon with $\Lambda\Lambda$-$N\Xi$-$\Sigma\Sigma$ 
coupled channel analysis is necessary.  Such a  direction is currently in
progress \cite{sasaki2010}. See also a related recent work on $H$-dibaryon in a different
 approach  \cite{Beane:2010hg}.

\section*{Acknowledgements}

I was  a Gerry Brown's posdoc at Stony Brook for two years starting from March 1988.
 Soon after I arrived at Stony Book, Gerry gave me a copy of his
 book ``The Nucleon-Nucleon Interaction". After more than 20 years since then,
 It is my great pleasure to  present some of our recent progresses 
  on lattice nuclear force in this volume to celebrate his 85$^{\rm th}$ birthday.
 This work was supported   
   by  Grant-in-Aid  for  Scientific Research  on  Innovative  Areas
   (No.~2004: 20105001,20105003). I thank the members of HAL QCD collaboration
   (S. Aoki, T. Doi, Y. Ikeda, T. Inoue, N. Ishii, K. Murano, H. Nemura, K. Sasaki)
   for fruitful discussions. Also, I thank Y. Fujiwara, C. Nakamoto,
   M. Oka, T. Takatsuka, R. Tamagaki and H. Toki for  stimulating
    discussions and encouragements.
 %


\end{document}